\documentclass{aa}

\usepackage{txfonts}
\usepackage{graphicx}

\newcommand{\lya}{Ly$\alpha$}
\newcommand{\lam}{$\lambda \lambda$}

\begin{document}

\title{The nature of the Ly$\alpha$-emission region of FDF-4691
       \thanks{Based on observations obtained at the ESO VLT
at Cerro Paranal, Chile.}}
\subtitle{}
\author{
C.~Tapken\inst{1}
        \and I.~Appenzeller\inst{1}  
        \and D.~Mehlert\inst{1}
        \and S.~Noll\inst{1}
\and S.~Richling\inst{2}
}
\institute{Landessternwarte Heidelberg-K\"onigstuhl,
            D-69117 Heidelberg, Germany
\and Institut d' Astrophysique de Paris, 98bis Bd Arago, 75014 Paris, France
}

\offprints{C. Tapken, Heidelberg (\email{ctapken@lsw.uni-heidelberg.de})}
\date{received; accepted}

\abstract{

In order to study the origin of the strong \lya\
emission of high-redshift starburst galaxies we observed and
modeled the emission of the $z = 3.304$ galaxy FDF-4691
(rest-frame $EW_{\rm{Ly}\alpha}$ = 103 \AA ). The observations show that
FDF-4691 is a young starburst galaxy with a (for this redshift)
typical metallicity. The broad, double-peaked profile of the
\lya\ emission line can be explained assuming a highly turbulent
emission region in the inner part of the starburst galaxy, and a surrounding 
extended shell of low-density neutral gas with a
normal dust/gas ratio and with Galactic dust properties.  The detection of the \lya\ 
emission line is explained by the intrinsic broad \lya\ emission and a low HI column 
density of the neutral shell. A low dust/gas ratio 
in the neutral shell is not needed to explain the strong \lya\ line.

}

\maketitle
\titlerunning{The nature of the \lya\ -emission region of FDF-4691}
\authorrunning{C. Tapken et al.}

\section{Introduction}
In all star-forming galaxies \lya\ photons are produced by recombination in 
HII regions ionized by young stars. However, these \lya\ photons are resonance scattered 
and thus have a large optical path in a neutral gas where they can be absorbed by dust grains
(see e.g. Neufeld \cite{neufeld}). 
This may explain the absence of \lya\ emission from  many local and  medium-redshift 
starburst galaxies (Charlot \& Fall \cite{charlot}).
On the other hand, \lya\ emission is often the most conspicuous feature of the spectra
of high-redshift galaxies. In fact, \lya\ emission is one of the most efficient tools to 
detect and identify high-redshift galaxies (e.g. Rhoads et al. \cite{rhoads}; Hu et al. \cite{hu}). 
Of 25 galaxies with $z > 5$ discussed by Taniguchi et al. (\cite{taniguchi}), all but two show strong 
\lya\ emission. Moreover, in short exposure spectra of high-redshift galaxies often 
no continuum is detected, while the \lya\ emission line is conspicuously present.
Kudritzki et al. (\cite{kudritzki}) argued that such galaxies must have a very low dust content. 
However, as noted by Kunth et al. (\cite{kunth}) a higher escape probability of \lya\ photons 
could also be caused by a suitable velocity field, which reduces the number of resonance scattering 
events.

In order to clarify the cause of the strong \lya\ emission of high-redshift galaxies, we 
observed the \lya\ emission line galaxy FDF-4691 (from the catalog of Heidt et al. \cite{heidt}) 
at low and medium spectral resolution. Moreover, we carried out radiative transfer model 
computations to reproduce the observed complex \lya\ line profile and to constrain the 
velocity fields and physical conditions of the \lya\ emitting volume.
Throughout this letter we adopt  $\Omega_{\Lambda}$ = 0.7, $\Omega_M$ = 0.3
and H$_{0}$ = 70 km s$^{-1}$Mpc$^{-1}$.

\section{Observations and data reduction}

\begin{figure}
\resizebox{\hsize}{!}{\includegraphics[angle=-90]{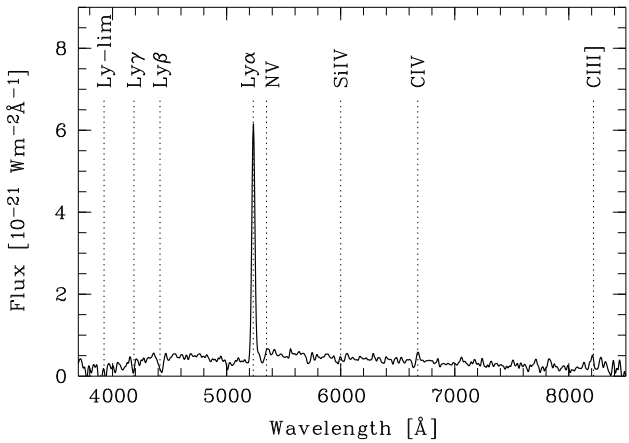}}
\caption{The low-resolution spectrum of FDF-4691. The dotted vertical lines indicate the
 positions of various expected lines.}
\label{4691lra} 
\end{figure}

\begin{figure}
\resizebox{\hsize}{!}{\includegraphics[angle=-90]{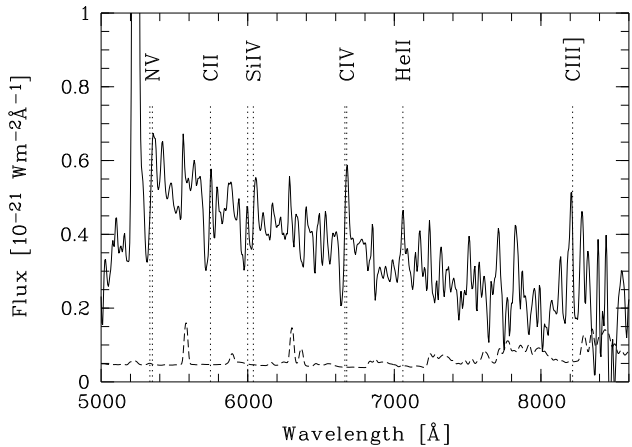}}
\caption{Scaled version of  Fig. 1. The dashed line indicates the noise level. }
\label{4691lrb}
\end{figure}

The low-resolution spectrum ($R = 200$) of FDF-4691 (Figs. \ref{4691lra} and \ref{4691lrb}) has been taken from 
the atlas of  Noll et al. (\cite{noll}), where the observational details are described. It is based 
on a total integration time of 550 min and has a continuum SNR of about 7 between 4000 and 8000 \AA .

Medium-resolution spectroscopic observations of FDF-4691 were obtained in August 2002 in service mode, 
together with other objects. 8 exposures with 47 min integration time each were taken using FORS2 and 
the 1400V grism resulting in a total integration time of 6.25 h. The co-added spectrum covers the range 
from 4500 to 6100 \AA ~and has a resolution of $R \approx$ 2000 ($\Delta v \approx$ 150 km s$^{-1}$). The spectra 
were reduced using MIDAS. Since FDF-4691 is relatively faint ($m_{\rm{I}} = 24.3$ mag), the continuum 
SNR of the medium resolution spectrum is $\le$ 3 throughout the observed spectral range. Hence this spectrum 
provides little additional information on the absorption features.

\begin{figure}
\resizebox{\hsize}{!}{\includegraphics[clip=true]{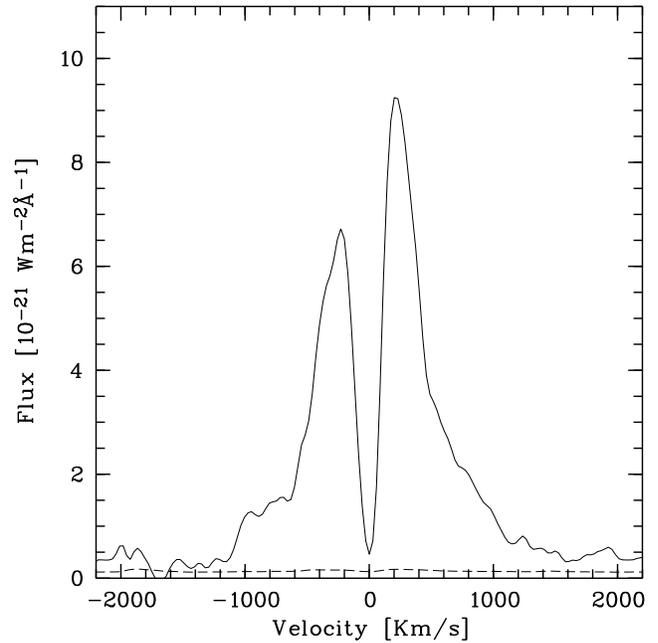}}
\caption {\lya\ line profile as derived from the medium-resolution spectrum of 
FDF-4691. The dashed line indicates the noise level.  }
\label{4691hra}  
\end{figure}

A much better SNR ($\le$ 60) was reached for the \lya\ emission line profile. Fig. \ref{4691hra} 
shows the observed profile. The abscissa gives the radial velocity relative to the central absorption
component (which corresponds to a redshift of $z = 3.304$).

\section{Results}

\subsection{Basic observed properties}
Apart from the exceptionally strong \lya\ line the low-resolution spectrum (Fig. \ref{4691lra}) shows a 
for this redshift  normal starburst spectrum. The continuum break at the wavelength of \lya\ is caused
by absorption of the \lya\ forest. 
The  UV luminosity at 1500 \AA ~is $L_{\rm{UV}}$ = 1.63 $\pm$ 0.06 $\times$ 10$^{34}$ W  
\AA$^{-1}$, which is typical for galaxies between z = 3 and z = 4 in the FDF spectroscopic survey 
(Noll et al. 2004).
The rest-frame \lya\  equivalent width is 103 $\pm$ 15 \AA , while the line flux amounts to 
$I_{\rm{Ly}\alpha} = 18.8 \pm 0.6 \times  10^{-20}$ W m$^{-2}$. These are only  lower limits 
since \lya\ photons are more affected by dust than the surrounding continuum. The Ly$\alpha$ luminosity is $L_{\rm{Ly}\alpha} 
= 1.83 \times  10^{36}$ W, if the emission is isotropic.  

As shown by Fig. \ref {4691lrb} at least three prominent stellar wind lines are detected:
NV \lam 1239, 1243 and CIV \lam 1548, 1551 show  P Cygni profiles and SiV \lam 1394, 1403 is mainly 
in absorption. HeII $\lambda$1640  (total flux 5.3 $\pm$ 1.6 $\times$  10$^{-21}$ W m$^{-2}$) and CIII] $\lambda$1909 
(total flux 8.8 $\pm$ 2.0 $\times$  10$^{-21}$ W m$^{-2}$) appear in emission. 
A strong absorption ($|EW| =  3.8 \pm 0.9$ $\AA$) blue-wards from the indicated position of CII $\lambda$1335 
is discussed in Section 3.3.

\subsection{AGN or starburst?}

Our \lya\ profile ($FWZI \approx 2000$ km s$^{-1}$) differs from \lya ~profiles of high-redshift
galaxies observed by, e.g., Dawson et al. (\cite{dawson}), which show  narrower, asymmetric profiles with 
$FWHM \approx 300$ km s$^{-1}$  with a sharp blue cut-off and a red wing. On the other hand, van 
Ojik et al. (\cite{ojik}) found for high-redshift radio galaxies symmetric \lya ~profiles with line 
widths of the order of $\approx 1500$ km s$^{-1}$. 
As some of these profiles have the same line widths and similar profiles as FDF-4691, one might suspect 
that the \lya\ line of FDF-4691 is excited by an AGN.

However, the hard radiation field of an AGN would be expected to result in NV / \lya , 
CIV / Ly$\alpha$ and CIII] $\lambda$1909  / Ly$\alpha$ line ratios which are not observed.
For example,  the observed CIV / \lya\ = 0.025  is ten times smaller than the 
ratio in the composite spectra of AGNs (CIV / \lya\ = 0.25, Osterbrock \cite{osterbrock}).
 
Moreover radio observations of the FDF have shown that at 1.4 GHz no radio source $>$ 100 $\mu$Jy 
is present at the position of FDF-4691 (Wagner 2003, private communication).

Therefore, we conclude that the \lya\ emission of FDF-4691 is excited by the continuum radiation of the hot stars
of this galaxy.  

\subsection{Comparsion with STARBURST99 models}
\begin{figure} 
\resizebox{\hsize}{!}{\includegraphics[angle=-90]{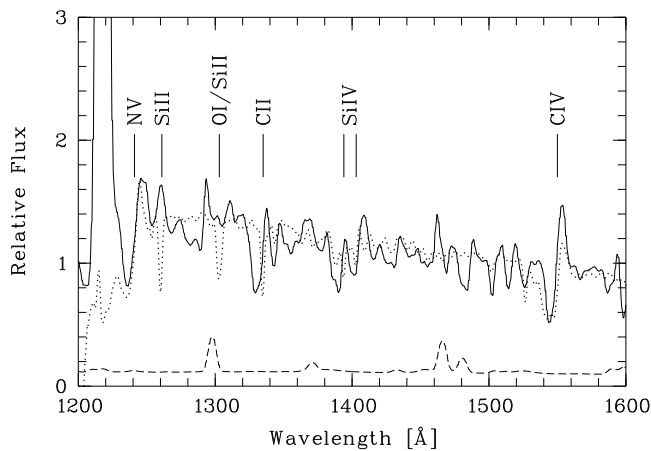}}
\caption{Low-resolution spectrum of FDF-4691 (solid line) and STARBURST99 spectrum (dotted line). 
The discrepancy at the interstellar absorption lines (SiII, OI/SiII and CII) is due to the fact that these are 
not modeled consistently with the extinction by STARBURST99. The dashed line indicates the noise level.}
\label{4691sb99}
\end{figure} 

In order to constrain the starburst age and metallicity of FDF-4691 we compared our spectrum with STARBURST99 
models (Leitherer et al. \cite{leitherer}), assuming a Calzetti et al. (\cite{calzetti}) reddening curve. The best 
fit for the UV slope and the profiles of the stellar wind lines was found for 
continuous star-formation, low starburst age ($<$ 20 Myr), LMC/SMC metallicity, a non-Salpeter IMF of $\alpha = 1.5$  
and $E(B-V) = 0.1$ mag (Fig. \ref{4691sb99}). 
The metallicity derived from the CIV absorption equivalent  
width ($|EW| = 3.0 \pm 0.9$ $\AA$) using the calibration given in Mehlert et al. (\cite{mehlert}) 
is also in agreement with the LMC/SMC value.
This metallicity is normal for this redshift (Mehlert et al. \cite{mehlert}).
The reddening of $E(B-V) = 0.1$ mag is among the lowest observed 
in this redshift range  in the FDF spectroscopic survey. Note that this value depends on the reddening curve used. 
The steeper extinction law of the SMC (Bouchet et al. \cite{bouchet}) would yield $E(B-V) = 0.03$ mag. 
The low reddening is not unexpected in view of the anticorrelation 
between the strength of the \lya\ line and the reddening as derived from the UV continuum slope $\beta$ 
found by Shapley et al. (\cite{shapley}). But deriving all the parameters mentioned above from UV data alone is not
very constraining. Other combinations also provide  good fits, including  models with an instantaneous
starburst and a very young age ($\simeq$ 5 Myr).

The CIV \lam 1548, 1551 emission component is stronger in FDF-4691 than in the model. This may indicate 
the presence of early O stars or Wolf-Rayet stars, which are known to show such CIV profiles (Wu et al. \cite{wu}; 
Walborn \& Panek \cite{walborn}; Willis \& Garmany \cite{willis}).  The fact that the low-ionisation interstellar 
lines (SiII $\lambda$1260, OI/SiII $\lambda$1303, CII $\lambda$1335), 
visible in the STARBURST99 model, are not present in FDF-4691, is again consistent with the correlation found 
by Shapley et al. (\cite{shapley}) between the reddening and the strength of the interstellar absorption lines. 
The presence of the strong  CII $\lambda$1335 line (blue-shifted by 1400 kms$^{-1}$ in
respect to the \lya\ line) may be explained by the presence of WC stars in FDF-4691, which tend
to show strong blue-shifted CII $\lambda$1335 lines (Willis \& Garmany \cite{willis}). WC stars may also provide an 
explanation for the strength of the HeII and CIII] emission. The presence of very young stars and WR stars 
together with the large equivalent width of \lya\, indicate that the starburst is very young ($< 5$ Myr).   
 
The star-formation rate of FDF-4691 was obtained from the \lya\ luminosity ($SFR_{\rm{Ly}\alpha} = 16.5 \pm 
 0.5$ M$_{\odot}$ yr$^{-1}$) and from the UV luminosity ($SFR_{\rm{UV}} = 
17.1 \pm 0.8$ M$_{\odot}$ yr$^{-1}$). In both cases we used the calibration given in Kennicutt (\cite{kennicutt}). 
The agreement indicates  that \lya\ is not significantly attenuated relative to the continuum.

\section{Comparison with line profile models}

\begin{figure}
\resizebox{\hsize}{!}{\includegraphics[clip=true]{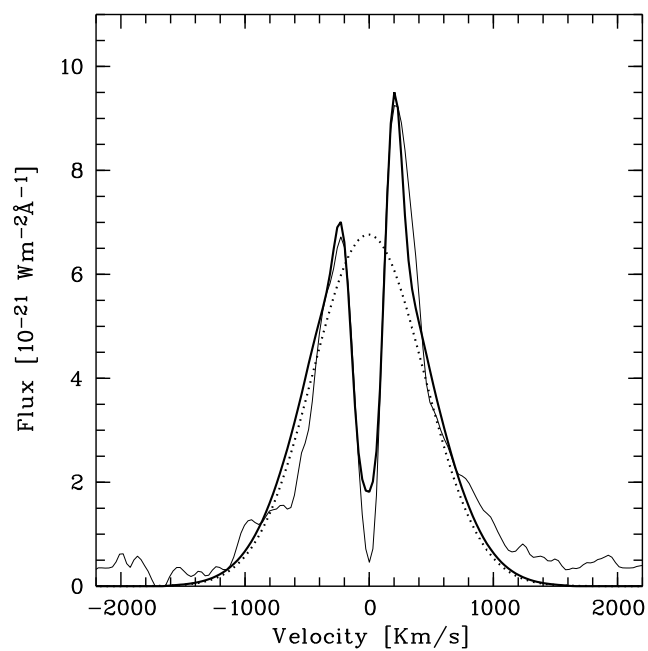}}
\caption{Comparsion of the \lya\ line of
FDF-4691 (thin line) and  the best-fit theoretical model (thick line). The intrinsic
emission profile is also shown (dotted line).}
\label{4691hrb}  
\end{figure}

In order to constrain the physical properties of the Ly$\alpha $
emitting volumes of FDF-4691 we calculated model profiles using 
the finite element line formation code of
Richling and Meink\"ohn (Richling et al. \cite{richling}; Meink\"ohn
\& Richling \cite{meinkoehn}), which is particularly well suited for
calculating the radiative transfer in a non-static scattering medium.
Since the width and the shape of the lower part of the observed line profile
is best explained by the emission of (high-velocity) turbulent
H II regions, while the narrow central absorption can be produced only in a
low-velocity H I layer in front of the emission region, we assumed
a spherical two-component model with a central line emission region
surrounded by a shell of neutral HI gas. Configurations of this type
have been suggested and investigated earlier 
by Tenorio-Tagle et al. (\cite{tenoriotagle}), Ahn et al. (\cite{ahna}, \cite{ahnb})
and Mas-Hesse et al. (\cite{mashesse}). For the emission from the central
region we assumed a broad Gaussian Ly$\alpha $ emission profile. This
assumption seems reasonable in view of the expected supernova rate and 
stellar wind activity in the observed compact starburst region. The turbulent velocity
of the central source was adjusted to fit the outer wings of the observed profile which are
not affected by the low-velocity gas of the shell. 

The computed model profile fitting the observed profile best is
reproduced in Fig. 5. The theoretical model was convolved with the 
instrumental profile. The parameters for this model are
turbulence velocities of about 600 km s$^{-1}$ in the emission
region, of 63 km s$^{-1}$ in the scattering shell,
and an outflow velocity of the shell of 12 km s$^{-1}$.
The central HI column density of the shell was $N(HI) \approx  
4 \times 10^{17}$ cm$^{-2}$, corresponding to an optical depth of 
the shell in the line center of $\tau_{0} \approx  5000$.

In the model described above the central absorption of the line profile
is produced by the removal of Ly$\alpha $ photons from
the line center by multiple resonance scattering. Without dust absorption 
the resonance scattering redistributes all these photons in
velocity space to produce the blue and and red peaks framing the 
central absorption. Calculations with  Galactic dust/gas ratio and Galactic 
dust properties and without dust in the neutral shell showed that
with the model parameters given above
the dust in the shell had no detectable effect on the line profile.
Dust in the central emission region, if not destroyed  
by the strong radiation field, only reduces the total emission without 
modifying the profile.

In this model the strong \lya\ emission line is caused by an intrinsic broad
\lya\ line, allowing a high fraction of \lya\ photons to escape unaffected by the
neutral shell, and a low column density of the neutral shell. Note that with 
Galactic dust/gas ratio the amount of dust in the shell is very low.

Since the shell turbulence velocity appears too high
with respect to the velocity of sound in (mainly) neutral 
interstellar matter and the rest intensity at the line 
center is somewhat to high in our model, we expect that we underestimated
the neutral column density. A higher neutral column density would increase
the separation between the two peaks, allowing to reproduce a similar model as in Fig. \ref{4691hrb} with
lower shell turbulence velocity.   Unfortunately,
at present, our code cannot handle much higher optical depths.

Additional model calculations with an increased dust/gas ratio and analytical 
models (Neufeld \cite{neufeld}) suggest that a up to  100 $\times$ higher column density of the neutral shell
with Galactic dust/gas ratio will not result in a significant reduction of \lya\ photons in the shell.
At higher neutral column densities the destruction of \lya\ photons will become important.

\section{Conclusion}

Our observations have shown that the Ly$\alpha$ galaxy FDF-4691
is a young starburst galaxy with a (for this redshift) normal
metallicity and a modest amount of reddening. The Ly$\alpha $ flux
of the galaxy appears not significantly more attenuated by
dust absorption than the UV continuum. According to our models
the observed line profile can be explained assuming that a 
turbulence-produced broad profile of a central emission region 
is modified by frequency redistribution in a resonance scattering
neutral shell around the central H II region. Although the
model fit does not constrain the physical parameters of the model
well (and the model may be much too simplistic), the computations
demonstrate that the detection of the high \lya\ flux is caused by a low neutral
column density and an  intrinsic broad emission line. A low dust/gas ratio 
in the neutral shell is not needed to explain the strong \lya\ line.

\begin{acknowledgements}
We are grateful to the referee for helpful comments.
Our research has been supported by the German Science Foundation DFG (SFB 439).
\end{acknowledgements}

\end{document}